\documentclass[conference]{IEEEtran}
\usepackage[T1]{fontenc}
\usepackage{mathtools} 
\usepackage{amsmath,amssymb,amsfonts}
\usepackage{cite}
\usepackage{graphicx}
\usepackage{psfrag}
\usepackage{subfigure}
\usepackage{bbding}
\usepackage{amsmath}
\usepackage{array}
\usepackage{algorithmicx}
\usepackage{algpseudocode}
\usepackage{setspace}
\usepackage{blindtext}
\usepackage{url}
\usepackage{epstopdf}
\usepackage{verbatim}
\usepackage{color}
\usepackage{amsmath}
\usepackage{bm}
\usepackage{multirow}
\usepackage[table,xcdraw]{xcolor}
\usepackage{booktabs}
\usepackage{makecell}
\usepackage{ntheorem}
\usepackage{textcomp}
\usepackage{amsmath}
\usepackage{float}
\usepackage{verbatim}
\usepackage{amssymb}

\newtheorem{Lem}{Lemma}

\newtheorem{Prob}{Problem}

\newcommand{\RNum}[1]{\uppercase\expandafter{\romannumeral #1\relax}}

\theoremseparator{:}

\title{A Variational Auto-Encoder Enabled Multi-Band Channel Prediction Scheme for Indoor Localization}
\author{Ruihao Yuan, Kaixuan Huang, Pan Yang, and Shunqing Zhang   \\
Shanghai Institute for Advanced Communication and Data Science, \\
Shanghai University, Shanghai, 200444, China\\
Email: \{yuan980728, xuan1999, yang\_pan, shunqing\}@shu.edu.cn}

\begin{document}
\maketitle

\begin{abstract}
Indoor localization is getting increasing demands for various cutting-edged technologies, like Virtual/Augmented reality and smart home. Traditional model-based localization suffers from significant computational overhead, so fingerprint localization is getting increasing attention, which needs lower computation cost after the fingerprint database is built. However, the accuracy of indoor localization is limited by the complicated indoor environment which brings the multipath signal refraction. In this paper, we provided a scheme to improve the accuracy of indoor fingerprint localization from the frequency domain by predicting the channel state information (CSI) values from another transmitting channel and spliced the multi-band information together to get more precise localization results. We tested our proposed scheme on COST 2100 simulation data and real time orthogonal frequency division multiplexing (OFDM) WiFi data collected from an office scenario. 
\end{abstract}

\section{Introduction} \label{sect:intro}

Indoor localization has received growing attention recently. Different from the outdoor localization and tracking tasks, the useful satellite signals in the outdoor environment are usually unreliable for many indoor applications due to the signal blockage. Even with many localization infrastructures available, the indoor localization tasks may suffer from the complicated multi-path signal refraction, reflection and blocking effects, while the localization accuracy is limited in general \cite{9585467}.

In order to improve the localization accuracy, the existing literature focuses on extending the range based \cite{4802191} or fingerprint based \cite{9685586} methods. For instance, a majorization minimization method using hybrid range based time-of-arrival (TOA) and received signal strength (RSS) information has been proposed in\cite{9723070}, where the proposed scheme can iteratively minimize the non-linear weighted least squares and the achievable localization accuracy can be improved to around 0.5 meter in terms of normalized mean square errors (NMSE). In the signal fingerprint based localization field, many augment and fusion frameworks have been developed as well\cite{Redzic2020ImageAW}, which cover a wide application of WiFi \cite{8423070}, ultra-wideband (UWB) \cite{8015125}, and visual images \cite{9001071}. As illustrated in \cite{9001071}, the indoor localization accuracy can be improved from 5.0 meters to around 0.3 meters by applying weighted access points (WAPs)-based WiFi matching, the
Gaussian weighted KNN (GW-KNN)-based image-level localization. An RSS-Image threshold-based fusion and particle filter fusion method learning algorithm has been proposed to incorporate multi-modal sensing data, which enables about 1 meter NMSE reduction\cite{Redzic2020ImageAW}.

Apart from the above fusion schemes, another effective method is to enlarge the observation windows, either in the time\cite{8862951} or spatial\cite{9685586} domain. 
In \cite{9685586}, the RMSE localization performance could be improved from 1.747 meters to 0.918 meters through multiple observations generated by dummy antennas. A natural extension is whether multi-band CSI samples are beneficial to improve the localization accuracy. Although the answer might be yes in a straight forward sense, there is quite limited literature to discuss the above problem due to the following reasons.

\begin{itemize}
    \item {\em Non-linear Cross-Band Correlation Characterization} In the conventional multi-band localization schemes, the achievable localization accuracy is in general limited using the auto-regression based schemes. This is due to the underlying assumption of linear cross-band correlations, as reported in \cite{Vasisht2016DecimeterLevelLW}. However, this assumption may not hold true in many practical systems \cite{shen2018joint}, and a detailed non-linear characterization is thus required. 
    \item {\em Backward Compatible with Plug-in Structure} In the practical implementation, the number of available localization bands might be different for many practical applications. Therefore, a more preferred scheme shall make it backward compatible to the conventional multi-band localization framework and a plug-in structure will be promising.
\end{itemize}

To address the above issues, we transform the original NMSE minimization problem into the equivalent evidence lower bound maximization problem by introducing some auxiliary variables. With this decomposed structure, we develop a variational auto-encoder (VAE) enabled multi-band channel prediction block to characterize the non-linear cross-band correlation, and plug in to the existing multi-band localization structure for high precision indoor localization. Through some numerical and prototype results, we show that our proposed scheme can achieve more than 20\% MSE improvement if compared with learning-based channel prediction or auto-regression based mechanisms.

The rest of the paper is organized as follows. In Section \ref{sect:sys}, we introduce the system  model and formulate the localization problem. The multi-band localization problem transform and the deep learning based solution are given in Section \ref{sect:prob}. We present the numerical and prototype experiment results in Section \ref{sect:exper} and make the conclusion in Section \ref{sect:conc}.

\section{System Model \& Problem Formulation}
\label{sect:sys}
In this section, we first introduce the mathematical models adopted in multi-band localization systems and then formulate the indoor localization problem in what follows.

\begin{figure}[htpb!]
    \centering
    \includegraphics[width=9cm]{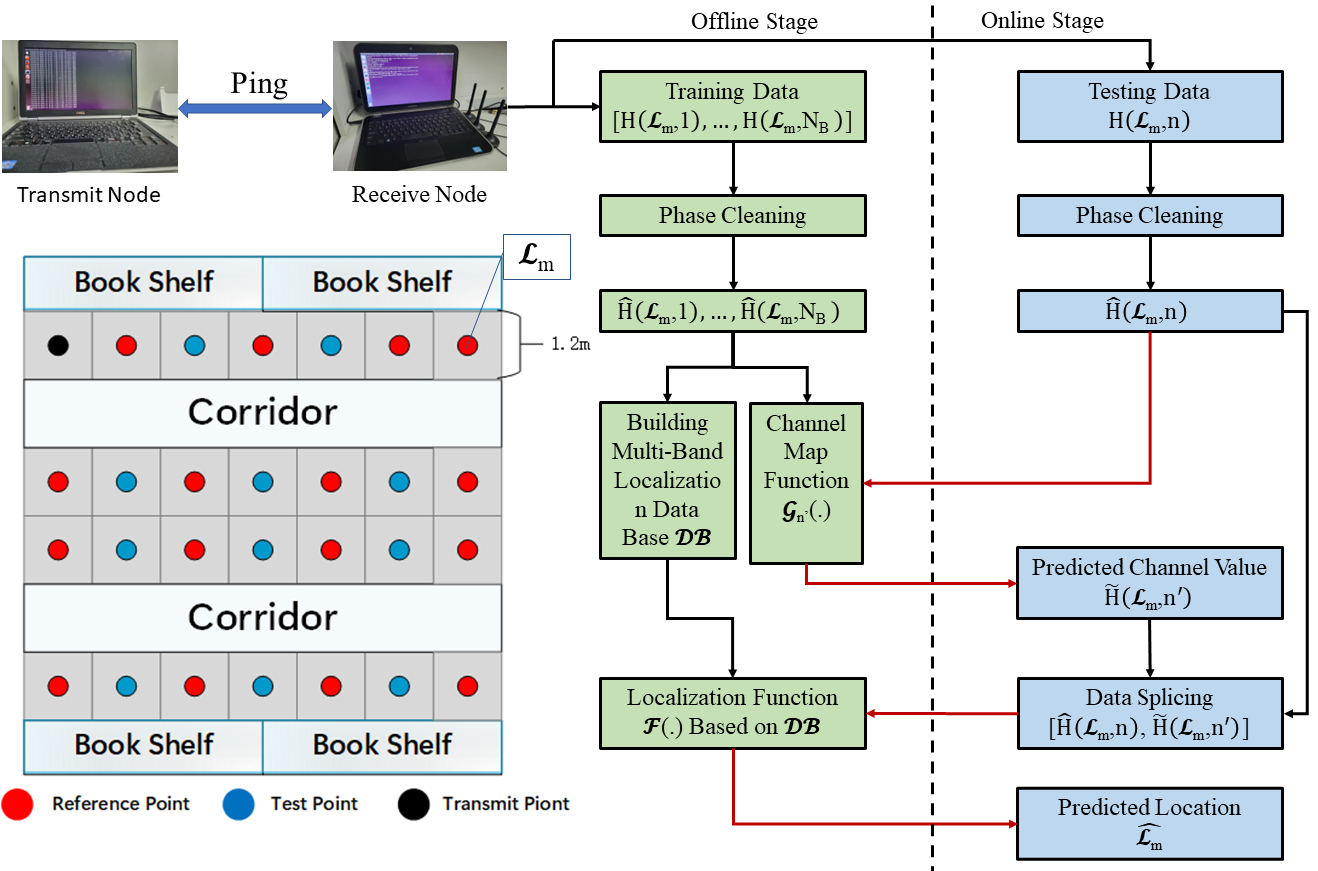}
    \caption{Structure of the whole system.}
    \label{fig:structure}
\end{figure}
Consider a multi-band orthogonal frequency division multiplexing (OFDM) enabled transmission system as shown in Fig.\ref{fig:structure}, and a single antenna localization entity is receiving WiFi signals from access points (AP) with $N_{T}$ transmit antennas. For any given location $\mathcal{L}$, the received signals $\mathbf{y}$ from the $i$-th antenna and the $n$-th frequency band, e.g., $\mathbf{y}^{i}(\mathcal{L},n) = [y^{i}_{1}(\mathcal{L},n), \ldots, y^{i}_{N_{sc}}(\mathcal{L},n)]$, are given by,
\begin{eqnarray}
\mathbf{y}^{i}(\mathcal{L},n) = \mathbf{H}^{i}(\mathcal{L},n) \mathbf{x}^{i}(\mathcal{L},n) + \mathbf{n}^{i}(\mathcal{L},n),
\end{eqnarray}
where $\mathbf{H}^{i}(\mathcal{L},n) \in \mathbb{C}^{N_{sc} \times N_{sc}}$, $\mathbf{x}^{i}(\mathcal{L},n), \mathbf{n}^{i}(\mathcal{L},n) \in \mathbb{C}^{N_{sc} \times 1}$ denote the channel fading coefficients, the transmitted symbols, and the additive white Gaussian noise with zero mean and unity variances, respectively. $N_{sc}$ is the number of sub-carriers per each frequency band, and $N_{B}$ represents the total number of frequency bands. According to the COST2100 channel model \cite{6393523}, the channel fading coefficients $\mathbf{H}^{i}(\mathcal{L},n)$ are given by,
\begin{eqnarray}
\mathbf{H}^{i}(\mathcal{L},n) & = & \sum_{p=1}^P \alpha_{p}(\mathcal{L},n) \cdot e^{-j\cdot2\pi\cdot n\cdot \tau_p(\mathcal{L})},
\end{eqnarray}
where $p$ is the index of fading paths, $P$ denotes the total number of multi-paths, and $\alpha_p(\mathcal{L},n)$ represents the path-loss coefficients of the $p$ path. In practice, $\alpha_p(\mathcal{L},n)$ can be affected by the locations of visible clusters, the direction-of-arrival and direction-of-departure of fading paths through the location $\mathcal{L}$, as well as different frequency responses through the band index $n$.

By estimating and collecting channel fading coefficients from different frequency bands together, we construct the localization database according to the following format.
\begin{eqnarray}
\mathcal{DB} = \left\{\left(\mathcal{L}, \hat{\mathbf{H}}(\mathcal{L},1), \ldots, \hat{\mathbf{H}}(\mathcal{L},N_B)\right)\right\},
\end{eqnarray}
where $\hat{\mathbf{H}}(\mathcal{L},n) = \{\hat{\mathbf{H}}^{i}(\mathcal{L},n)\}, \forall n \in [1, \ldots, N_B]$, denotes the measured channel responses of the $n$-th frequency band, after removing the random phase offset as explained in \cite{7417517}.

With the established database $\mathcal{DB}$, our proposed localization system shall identify the location $\mathcal{L}_{m}$ from the real time measured channel responses $\hat{\mathbf{H}}(\mathcal{L}_{m},n)$. Mathematically, the MSE minimization problem is given as follows.
\begin{Prob}[\em MSE Minimization] \label{prob:MSE}
The localization MSE minimization problem for our proposed localization system is given as follows,
\begin{eqnarray}
\underset{\mathcal{F}(\cdot)}{\textrm{minimize}} && \frac{1}{M}\sum_{m=1}^{M} \|\hat{\mathcal{L}}_{m} - \mathcal{L}_{m}\|^2_2, \\
\label{eqn:mini} \nonumber\\
\textrm{subject to} && \hat{\mathcal{L}}_{m} = \mathcal{F}\left(\mathcal{DB},\hat{\mathbf{H}}(\mathcal{L}_{m},n)\right), \forall m, \\
&& \hat{\mathcal{L}}_{m}, {\mathcal{L}}_{m} \in \mathcal{A},
\end{eqnarray}
 where $\mathcal{A}$ represents the feasible indoor localization areas, and $M$ denotes the total number of localization tasks and  $\mathcal{F}(\cdot)$ denotes the localization function.
\end{Prob}
The above problem is in general difficult to solve, since the optimal localization function $\mathcal{F}^{\star}(\cdot)$ can hardly be obtained by searching all the possible functions.

\begin{figure*}[h]
    \centering
    \includegraphics[width=7 in]{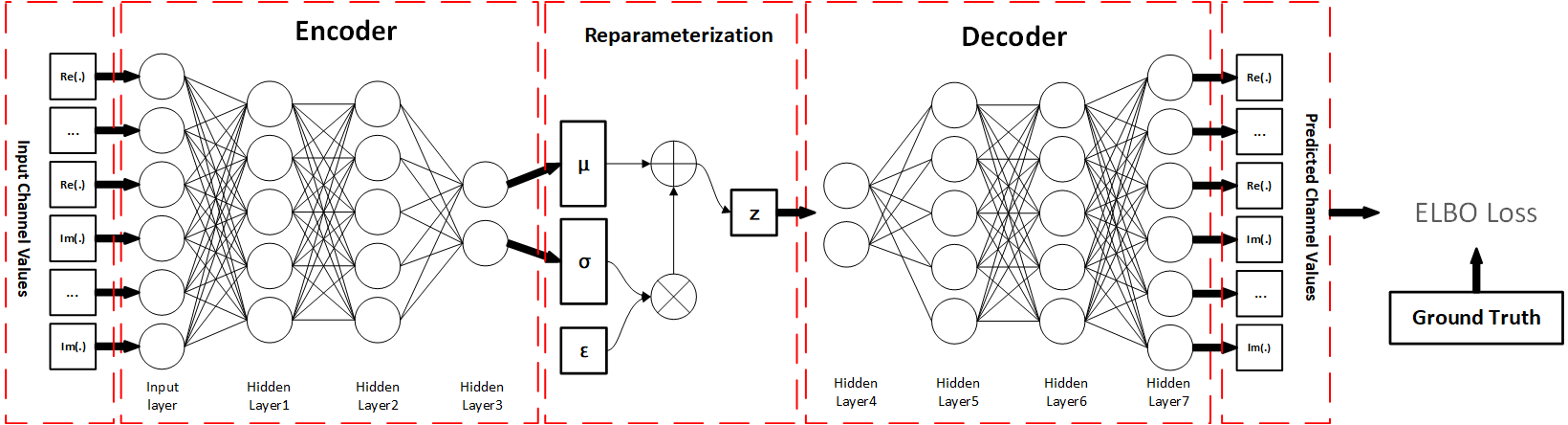}
    \caption{The structure of the proposed VAE model.}
    \label{fig:VAE structure}
\end{figure*}

\section{Proposed VAE Enabled Localization Scheme} 
\label{sect:prob}

In this section, we transform the above MSE minimization problem into the equivalent channel prediction error minimization problem, and propose the VAE enabled localization scheme in what follows. 

By introducing the auxiliary variables, $\{\tilde{\mathbf{H}}(\mathcal{L}_{m},n)\}$, the original MSE minimization problem can be transformed into the following format.
\begin{eqnarray}
\underset{\tilde{\mathcal{F}}(\cdot), \{\mathcal{G}_{n'}(\cdot)\}}{\textrm{minimize}} && \frac{1}{M}\sum_{m=1}^{M} \|\hat{\mathcal{L}}_{m} - \mathcal{L}_{m}\|^2_2, \\
\label{eqn:mini2} \nonumber\\
\textrm{subject to} && \hat{\mathcal{L}}_{m} = \tilde{\mathcal{F}}\left(\mathcal{DB},\tilde{\mathbf{H}}(\mathcal{L}_{m},1), \ldots, \tilde{\mathbf{H}}(\mathcal{L}_{m},N_B)\right), \nonumber \\ 
&& \tilde{\mathbf{H}}(\mathcal{L}_{m}, n') = \mathcal{G}_{n'} \left(\hat{\mathbf{H}}(\mathcal{L}_{m},n)\right), \\
&& \hat{\mathcal{L}}_{m}, {\mathcal{L}}_{m} \in \mathcal{A}, \forall m \in [1, M], \forall n' \in [1, N_B],
\end{eqnarray}
where $\{\mathcal{G}_{n'}(\cdot)\}$ denote the channel prediction functions from the current estimated band $\hat{\mathbf{H}}(\mathcal{L}_{m},n)$ to all $N_B$ frequency bands. With all the available predicted channel states of $N_B$ frequency bands, the optimal localization function  $\tilde{\mathcal{F}}^{\star}(\cdot)$ can be solved by standard machine learning technique as elaborated in \cite{8862951}. Specifically, we can apply deep neural network architecture with three hidden layers to model this non-linear relationship. By adopting the above optimized localization function $\tilde{\mathcal{F}}^{\star}(\cdot)$, the optimal channel prediction functions, $\{\mathcal{G}^{\star}_{n'}(\cdot)\}$, can be solved through the following posterior probability maximization problem.


\begin{Prob}[\em Evidence Lower Bound (ELBO) Maximization] \label{prob:ppm}
The ELBO maximization problem can be expressed as,
\begin{eqnarray}
\underset{\{\mathcal{G}_{n'}(\cdot)\}}{\textrm{maximize}} && \Psi(\tilde{\mathbf{H}}(\mathcal{L}_{m}, n'),\hat{\mathbf{H}}(\mathcal{L}_{m},n)),\\
\textrm{subject to} && \tilde{\mathbf{H}}(\mathcal{L}_{m}, n') = \mathcal{G}_{n'} \left(\hat{\mathbf{H}}(\mathcal{L}_{m},n)\right), \\
&& \forall n' \in [1, N_B],
\end{eqnarray}
where $\Psi(\tilde{\mathbf{H}}_1,\hat{\mathbf{H}}_2) = \mathbb{E}[\log P_r(\tilde{\mathbf{H}}_1|e(\hat{\mathbf{H}}_2))]  - \mathbb{D}_{KL}[P_r($ $e(\hat{\mathbf{H}}_2)|\tilde{\mathbf{H}}_1)\|P_r(e(\hat{\mathbf{H}}_2))]$ denotes the ELBO function as defined in \cite{kingma2013auto}. In the above expression, $\mathbb{E}(\cdot)$ represents the mathematical expectation operation. $P_r(A|B)$ denotes the probability distribution of the random variable $A$ condition on the distribution of $B$. $\mathbb{D}_{KL}[A\|B]$ denotes the KL-divergence between the probability distributions of $A$ and $B$. $e(\cdot)$ is a mapping function transforming the inner distribution into a lower dimension.
\end{Prob}

\begin{Lem} \label{lem:equ}
If $e(\hat{\mathbf{H}}_2)$ is a standard Gaussian distribution with zero mean and unit variance, and $P_r(\tilde{\mathbf{H}}_1|e(\hat{\mathbf{H}}_2))$ and $P_r(e(\hat{\mathbf{H}}_2)|\tilde{\mathbf{H}}_1)$ are assumed to be Gaussian, Problem~\ref{prob:MSE} and Problem~\ref{prob:ppm} are equivalent in terms of the Cramer-Rao Lower Bound (CRLB). 
\begin{IEEEproof}
Please refer to Appendix~\ref{appen:elbo} for the proof.
\end{IEEEproof}
\end{Lem}


\subsection{Network Structure Design}

Following our previous work \cite{7959171}, we adopt a multi-layer perceptron (MLP) neural network architecture with three hidden layers to approximate the function $\tilde{\mathcal{F}}^{\star}(\cdot)$. Rectified linear unit (ReLU) is chosen to be the activation function, and the dropout technique is applied to address the over-fitting issue. Detailed network parameters are listed in Table~\ref{tab:hyperpara}.

As illustrated in\cite{Alrabeiah2019DeepLF}, the mapping function $\mathcal{G}_{n'}(\cdot)$ exists and could not be easily characterized by any closed-form expression. To make it tractable to processing, we adopt the VAE structure as suggested in \cite{liu2021fire} to approximate the mapping function $\mathcal{G}_{n'}(\cdot)$ which projects the measured channel responses $\hat{\mathbf{H}}(\mathcal{L}_{m},n)$ into a lower dimensional space $e(\hat{\mathbf{H}}(\mathcal{L}_{m},n))$ via the encoder, and expands to the predicted channel responses $\tilde{\mathbf{H}}(\mathcal{L}_{m}, n')$ through the decoder. In VAE, the mentioned lower dimensional space is denoted by $z$. We use fully connected (FC) layers to model the encoder and decoder, respectively, and the detailed network structure is depicted in Fig. \ref{fig:VAE structure}. $Re(\cdot)$ and $Im(\cdot)$ are the real and imaginary part of the data, respectively.
In Table \ref{tab:VAEparam}, we list the network parameters with different sizes and all the activation function of different layers are chosen to be LeakyReLU\cite{maas2013rectifier}, in order to avoid dying ReLU problem compared with ReLU function. In addition, the dimension of $e(\hat{\mathbf{H}}(\mathcal{L}_{m},n))$ is selected to be 25 in the numerical evaluation.
 \begin{table} [ht]
\centering
\caption{The parameters of proposed VAE networks.}
\label{tab:VAEparam}
\footnotesize
\setlength{\tabcolsep}{6mm}
\renewcommand\arraystretch{1.5}
	\begin{tabular}{c|cc}  
    \hline
    \textbf{Structure}&\textbf{Layers}&\textbf{Size} \\
    \cline{1-3}
    \multirow{4}*{\textbf{Encoder}}& Input Layer&FC $2\times N_{sc}\times i$ \\
    \cline{2-3}
    ~&Hidden Layer 1&FC 64 \\
    \cline{2-3}
    ~&Hidden Layer 2&FC 64 \\
    \cline{2-3}
   ~&Hidden Layer 3&FC 100\\
    \cline{1-3}
    \multirow{4}*{\textbf{Decoder}}& Hidden Layer 4&FC 50\\
    \cline{2-3}
    ~&Hidden Layer 5&FC 64\\
    \cline{2-3}
    ~&Hidden Layer 6&FC 64 \\
    \cline{2-3}
   ~&Output Layer&FC $2\times N_{sc}\times i$\\
    \cline{1-3}
	\end{tabular}
\end{table}
\subsection{Loss Function Design}
\label{sec:loss}


In order to maximize the $ELBO$, we can intuitively maximize $\mathbb{E}[\log P_r(\tilde{\mathbf{H}}(\mathcal{L}_{m}, n')|z)]$ and minimize $\mathbb{D}_{KL}[P_r(z|\tilde{\mathbf{H}}(\mathcal{L}_{m}, n')\|P_r(z)]$ simultaneously. Without loss of generality, we assume $P_r(\tilde{\mathbf{H}}(\mathcal{L}_{m}, n')|z)$ and $Pr(z|\tilde{\mathbf{H}}(\mathcal{L}_{m}, n'))$ follow the Gaussian distribution with means $\mu$, $\mu'$ and variances $\sigma$, $\sigma'$, respectively\cite{kingma2013auto}. Therefore, the two terms in $ELBO$ are given by,
\begin{eqnarray}
\log P_r(\tilde{\mathbf{H}}(\mathcal{L}_{m}, n')|z) \sim -\frac{1}{2}\|\tilde{\mathbf{H}}(\mathcal{L}_{m}, n')-\hat{\mathbf{H}}(\mathcal{L}_{m}, n')\|^2_2
\end{eqnarray}
\begin{eqnarray}
\mathbb{D}_{KL}[P_r(z|\tilde{\mathbf{H}}(\mathcal{L}_{m}, n')\|P_r(z)] \nonumber\\= \mathbb{D}_{KL}(N(\mu',\sigma'^2)\|N(0,1)) \nonumber\\
= \frac{1}{2}(-\log \sigma'^2 + \mu'^2 + \sigma'^2 -1)
\end{eqnarray}

With the above understanding, we choose the loss function of VAE to be:
\begin{eqnarray}
\ell_{\mathcal{G}_{n'}(\cdot)} = \frac{1}{M}\sum_{m=1}^{M}\|\tilde{\mathbf{H}}(\mathcal{L}_{m}, n')-\hat{\mathbf{H}}(\mathcal{L}_{m}, n')\|^2_2 \nonumber\\ + \beta\times\frac{1}{2}(-\log \sigma'^2 + \mu'^2 + \sigma'^2 -1)
\end{eqnarray}
To balance the contribution of the two reconstruct loss and KL-divergence in $ELBO$, the $\beta$-VAE is proposed in \cite{higgins2016beta}, it balances the two parts of loss function during training by a hyperparameter $\beta$. The hyperparameters of the above networks in the training process are shown in Table.~\ref{tab:hyperpara}. 
\begin{table} [ht]
\centering
\caption{The hyperparameters of proposed networks.}
\label{tab:hyperpara}
\footnotesize
\setlength{\tabcolsep}{6mm}
\renewcommand\arraystretch{1.5}
	\begin{tabular}{ccc}  
    \hline
    \textbf{Parameter}&\textbf{VAE value}&\textbf{DNN value} \\
    \cline{1-3}
    learning rate& $10^{-5}$&$10^{-6}$\\
    \cline{1-3}
    optimizer&Adam&Adam\\
    \cline{1-3}
    epoch&50&90\\
    \cline{1-3}
    $\beta$&$10^{-1}$&-\\
    \cline{1-3}
	\end{tabular}
\end{table}
\section{Numerical and Prototype Results}
\label{sect:exper}
In this section, we compare the proposed VAE enabled multi-band indoor localization scheme with two conventional baseline methods. {\em Baseline 1: Learning-based Channel Prediction \cite{alrabeiah2019deep,huang2019deep,yang2019deep}}, which applies MLP to learn the cross-band correlations. {\em Baseline 2: Auto-regression with Extended Kalman Filter (EKF) \cite{shen2018joint}}, which relies on EKF with iterative detector decoder for the channel estimation and prediction. {\em Baseline 3: Real-time Sampled Data,} we directly collect all the channel responses from different bands simultaneously and perform the localization using the conventional multi-band localization scheme. Both numerical and prototype experiments are performed on 5 GHz WiFi scenario with $N_B=3$ bands, where the corresponding center frequencies are given by 5.765 GHz, 5.785 GHz, and 5.805 GHz (Band index 153, 157, and 161), respectively. Each band contains 20 MHz with 64 OFDM sub-carriers. Before prototype experiments, we tested our system on simulation numerical experiments, and the CSIs in numerical evaluations are generated according to the COST2100 model \cite{6393523} and the CSIs in prototype evaluations are collected from two laptops equipped with Intel 5300 network interface cards, where one laptop is running the access point mode and the other is the localization entity as shown in Fig.\ref{fig:structure}. 
Other simulation and experimental parameters are listed in Table \ref{tab:deployment}. 
\begin{table} [ht]
\centering
\caption{The parameter of numerical and prototype experiments.}
\label{tab:deployment}
\footnotesize
\setlength{\tabcolsep}{4mm}
\renewcommand\arraystretch{1.5}
	\begin{tabular}{cccc}  
    \hline
    \textbf{Parameter}&\textbf{Value}&\textbf{Parameter}&\textbf{Value} \\
    \cline{1-4}
    AP & 1& Training data-set&$10000\times 16$\\
    \cline{1-4}
    RP& 16& Testing data-set&$2000\times 11$\\
    \cline{1-4}
    TP&11&$N_B$& 3\\
    \cline{1-4}

	\end{tabular}
\end{table}

\subsection{Numerical Results}

\begin{figure}[htbp]
\centering
     \includegraphics[width=3.0 in]{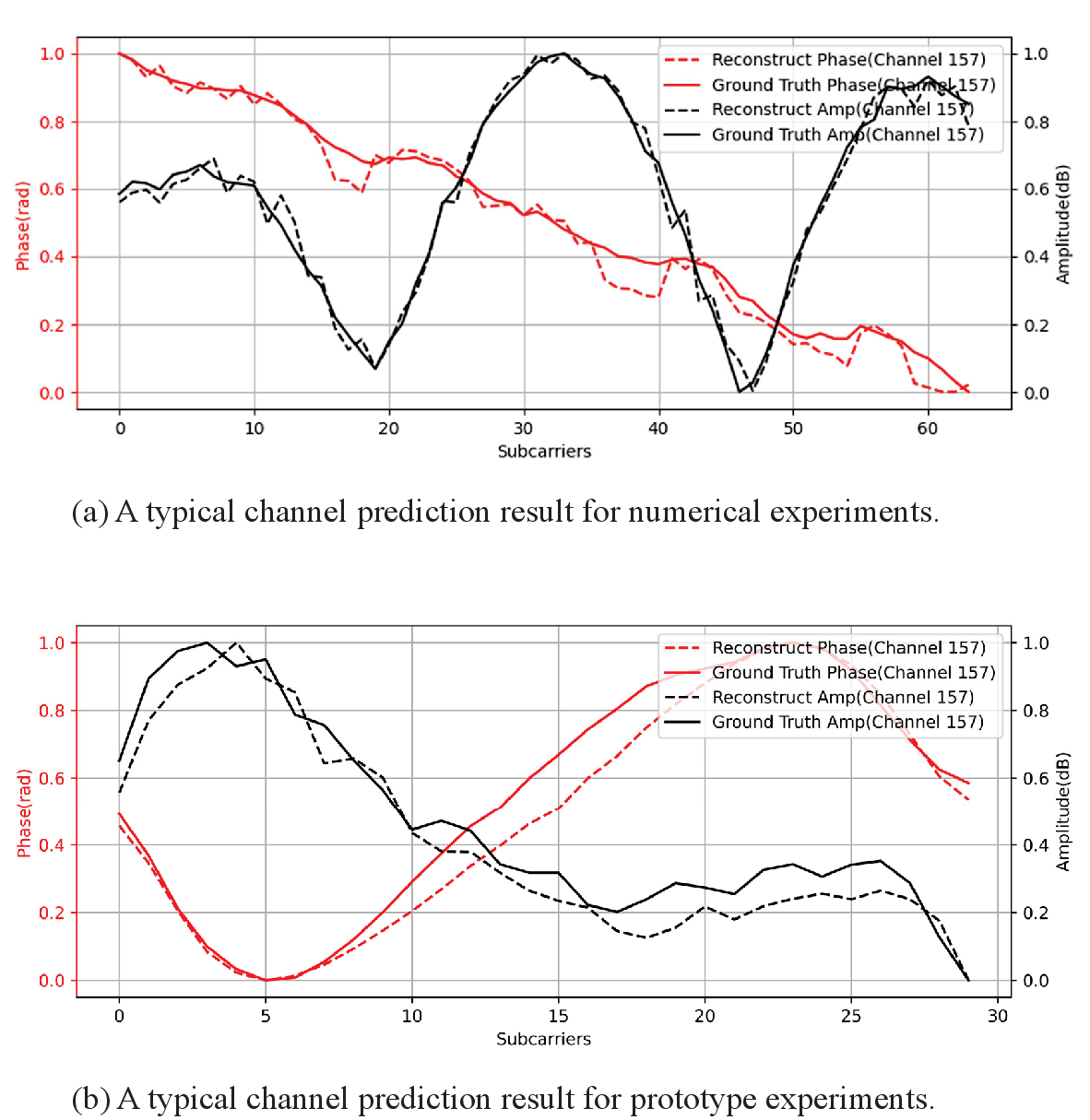}
    
    \caption{The amplitude and phase values of predicted channel in both numerical and prototype experiments.}
    \label{all abs and phase}
\end{figure}

In the first experiment, we numerically plot the estimated channel responses and compare with the predicted value from the VAE network in Fig.~\ref{all abs and phase}, where black and red lines denote the amplitude and phase information, respectively. As shown in Fig. \ref{all abs and phase}(a), the VAE predicted results (dashed lines) and the estimated channel responses (solid lines) are matched quite well.

\begin{figure}
    \centering
    \includegraphics[width=9.0 cm]{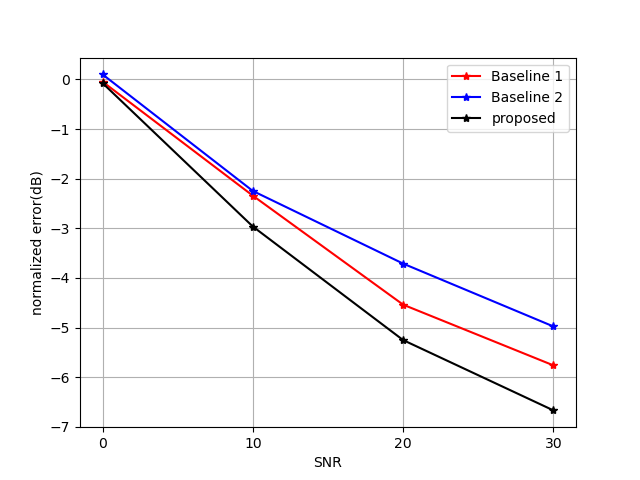}
    \caption{The channel coefficient normalized error of different methods of channel prediction.}
    \label{fig:prediction results}
\end{figure}

In the second experiment, we compare the channel prediction accuracy of different channel prediction schemes using the channel coefficient normalized error  (CCNE) performance \cite{Alrabeiah2019DeepLF} defined as,
\begin{eqnarray}
\label{NE}
CCNE = 10\lg{\left(\frac{\|\tilde{\mathbf{H}}(\mathcal{L}_{m}, n')-\hat{\mathbf{H}}(\mathcal{L}_{m}, n')\|^2}{\|\hat{\mathbf{H}}(\mathcal{L}_{m}, n')\|^2}\right)}. 
\end{eqnarray}
As shown in Fig.\ref{fig:prediction results}, the proposed VAE based channel prediction scheme can achieve about 54\% and 41\% CCNE improvement\footnote{Since {\em Baseline 3} is the real-time sampled channel responses, we do not plot the CCNE result for this case.} for $SNR = 30$ dB, if compared with {\em Baseline 1} and {\em Baseline 2}, respectively. 

\begin{figure}
    \centering
    \includegraphics[width=9cm]{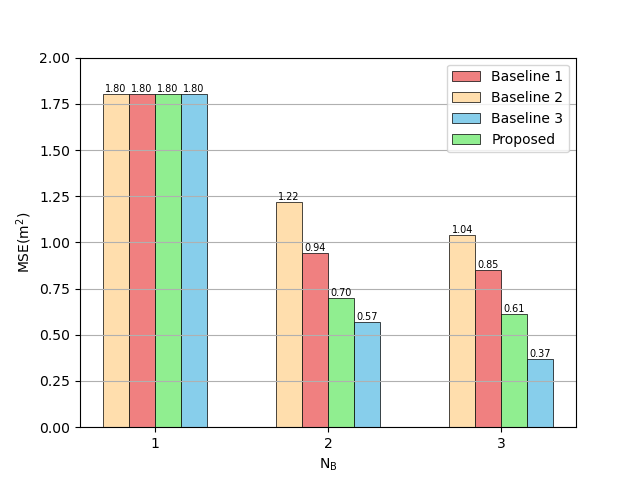}
    \caption{The localization results for numerical experiments}
    \label{fig:localization error}
\end{figure}

In the third experiment, we compare the MSE performance of different localization schemes, where the simulation results are shown in Fig.\ref{fig:localization error}. From this figure, we can observe that the proposed VAE enabled multi-band cooperative localization scheme is able to significantly outperform {\em Baseline 2} by at least 24\% MSE improvement, and achieve the upper bound of multi-band cooperative localization scheme with real-time measured responses ({\em Baseline 3}).

\subsection{Prototype Results}

In this part, we redo the above three experiments using our prototype localization systems, where the topology is shown in Fig.\ref{fig:structure}. Before passing to the multi-band localization system, we eliminate the carrier frequency offset (CFO) and sampling frequency offset (SFO) as introduced in \cite{7417517} to obtain more reliable channel responses. As expected, we can observe a close match between the estimated and predicted channel responses in the first experiment. In the second experiment, the CCNE values for {\em Baseline 1}, {\em Baseline 2}, and the proposed schemes are -1.6417 dB, 0.5641 dB, and -2.2234 dB, respectively.

\begin{figure}
    \centering
    \includegraphics[width=9cm]{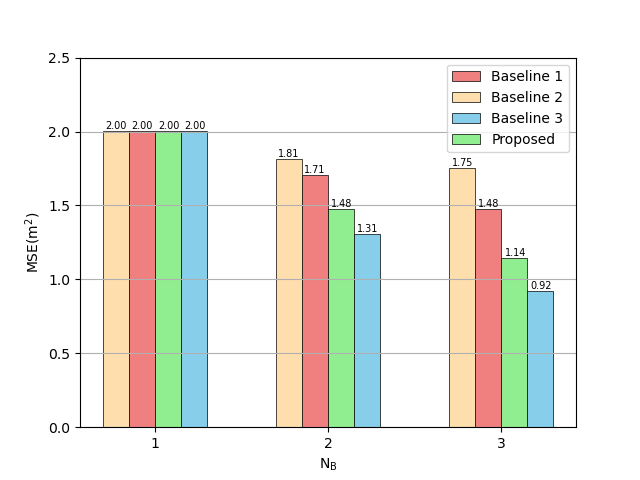}
    \caption{The localization results for prototype experiments}
    \label{real_data results}
\end{figure}

In the third experiment, the achieved MSEs of different multi-band localization schemes are shown in Fig.~\ref{real_data results}. Compared with {\em Baseline 1} and {\em Baseline 2}, our proposed VAE enabled multi-band localization scheme can achieve 47\% and 13\% improvement, respectively. Although the achievable localization performance improvement is slightly reduced if compared to numerical simulations, we can still show the effectiveness of the proposed VAE enabled multi-band localization schemes. 

\section{Conclusion} \label{sect:conc}
In this paper, we propose a novel VAE enabled multi-band indoor localization scheme. Different from conventional approaches, we apply VAE structure to describe the non-linear cross-band correlations, and expand the single band measured channel response to multiple bands via channel prediction. By incorporating measured and predicted channel responses of multiple bands, we can achieve 13\% to 54\% MSE improvement in the numerical and prototype experiments, if compared with other traditional multi-band localization schemes.

\appendices
\section{Proof of Lemma \ref{lem:equ}}
\label{appen:elbo}
The equivalent channel prediction posterior probability maximum problem for indoor localization is given as follows,
\begin{eqnarray}
{\textrm{maximize}} && \sum_{m=1}^{M} \log P_r(\tilde{\mathbf{H}}(\mathcal{L}_{m}, n')) 
\end{eqnarray}
 However, this problem is an intractable problem, so we transform this posterior probability maximization into the following evidence lower bound (ELBO) maximization problem.

\begin{eqnarray}
&&\log P_r(\tilde{\mathbf{H}}(\mathcal{L}_{m}, n')) \nonumber \\
&&= \log P_r(\tilde{\mathbf{H}}(\mathcal{L}_{m}, n'),e(\hat{\mathbf{H}}(\mathcal{L}_{m},n))) \nonumber\\
&&- \log P_r(e(\hat{\mathbf{H}}(\mathcal{L}_{m},n))|\tilde{\mathbf{H}}(\mathcal{L}_{m}, n'))\nonumber \\ 
&&= \mathbb{E}[\log (P_r(\tilde{\mathbf{H}}(\mathcal{L}_{m}, n'),e(\hat{\mathbf{H}}(\mathcal{L}_{m},n)))/P_r(e(\hat{\mathbf{H}}(\mathcal{L}_{m},n))))   \nonumber\\ &&-\log (P_r(e(\hat{\mathbf{H}}(\mathcal{L}_{m},n))|\tilde{\mathbf{H}}(\mathcal{L}_{m}, n'))/P_r(e(\hat{\mathbf{H}}(\mathcal{L}_{m},n))))]\nonumber\\ 
&&= ELBO \nonumber\\
&&+ D_{KL}(P_r(e(\hat{\mathbf{H}}(\mathcal{L}_{m},n))|\tilde{\mathbf{H}}(\mathcal{L}_{m}, n'))\nonumber\\
&&\|P_r(e(\hat{\mathbf{H}}(\mathcal{L}_{m},n))|\tilde{\mathbf{H}}(\mathcal{L}_{m}, n'))) \nonumber \\
&&\ge ELBO \label{eqn:ELOB}
\end{eqnarray}
In order to reflect the relationship between the independent variables, we denote the loss function  $ELBO$ as $\Psi(\tilde{\mathbf{H}}(\mathcal{L}_{m}, n'),\hat{\mathbf{H}}(\mathcal{L}_{m},n))$. The maximization problem of posterior probability problem could be converted to the maximization problem of $ELBO$.

According to \cite{4802191}, the Cramer-Rao Bound(CRLB) of CSI ranging error is formulated as:
\begin{eqnarray}
CRLB = \frac{c^2}{8\pi^2\beta^2SNR} \label{CRLB_one}
\end{eqnarray}
$c$ represents the speed of light, $SNR = E_p/N_0$, where $E_p$ is the average received energy, and $N_0$ is the energy of received noise, and $\beta$ is the transmit effective bandwidth, which means the CRLB of localization error decreases as the transmission bandwidth increases, so we choose to splice CSI data on two channel bands to improve the localization accuracy.

The effective bandwidth on the predicted channel and the CRLB for the spliced channel is given by:
\begin{eqnarray}
&& \beta' =P_r(\tilde{\mathbf{H}}(\mathcal{L}_{m}, n'))\beta \\
&&\hat{\beta} = \beta + \beta'\\
&&\widehat{CRLB} = \frac{c^2}{8\pi^2\hat{\beta}^2SNR} \le CRLB
\end{eqnarray}
$\beta'$ is the effective bandwidth of the predicted channel, $\hat{\beta}$ is the total effective bandwidth of the spliced channel. $\widehat{CRLB}$ is the range error lower bound of the spliced channel which is less than CRLB of single channel, getting a more accuracy prediction could extend the efficient bandwidth, this is the reason for Problem~\ref{prob:MSE} and Problem~\ref{prob:ppm} are equivalent. 

\bibliographystyle{IEEEtran}
\bibliography{IEEEabrv,bb_rf}

\end{document}